\documentclass[fleqn,usenatbib]{mnras}

\usepackage{newtxtext,newtxmath}

\usepackage[T1]{fontenc}

\DeclareRobustCommand{\VAN}[3]{#2}
\let\VANthebibliography\thebibliography
\def\thebibliography{\DeclareRobustCommand{\VAN}[3]{##3}\VANthebibliography}

\usepackage{graphicx}	
\usepackage{amsmath}	
\usepackage{hyperref}
\usepackage{ulem}

\title[Ten(+) Years of Binaries in Gaia]{Nominal thresholds for good astrometric fits, and prospects for binary detectability, for the full extended \textit{Gaia} mission}

\author[F. Guerriero et al.]{
F. Guerriero$^{1}$\thanks{\href{mailto:guerriero533@gmail.com}{guerriero533@gmail.com}}, 
Z. Penoyre$^{1}$, 
A. G. A. Brown$^{1}$
\\
$^{1}$Leiden Observatory, Leiden University, Niels Bohrweg 2, 2333 CA Leiden, The Netherlands
}

\pubyear{\the\year{}}

\begin{document}
\label{firstpage}
\pagerange{\pageref{firstpage}--\pageref{lastpage}}
\maketitle

\begin{abstract}
The full extended \textit{Gaia} mission spans slightly over 10 years, whilst the current data releases represent only a fraction of that timescale, 34 months in Gaia's third data release (DR3). A longer time baseline improves astrometric fits by lowering the noise and making consistently bad fits (for example, due to binarity) more apparent. In this paper, we use simulated binaries from the \textit{Gaia Universe Model} to examine the long-term astrometric behaviour of single stars and stellar binaries. We calculate nominal upper limits on the spread of goodness of astrometric fits for well-behaved single stars. Specifically, for the RUWE parameter, we predict thresholds of DR4 ($\rm RUWE_{lim}=1.15$) and DR5 ($\rm RUWE_{lim}=1.11$), based on the full mission nominal scanning law. These limits help identify poor astrometric fits and flag potential binary systems. We find that the number of detectable short-period binaries increases by 5-10\% per new data release, suggesting detections may be possible for orbital periods down to days. The number of detectable long-period systems increases by 10-20\%, with periods up to 100 years causing significant deviations in low and moderate-eccentricity binaries. Very eccentric systems with much longer periods (thousands of years) can still be detected if they pass through periapse during the observing window. The detectability of most systems is unaffected by the light ratio, although it is reduced for twin binaries. This is because astrometric motion is primarily a dynamical effect and depends mainly on the mass ratio, causing twin binaries to remain largely undetected. By contrast, the extended time baseline significantly enhances the detected binary population across the main-sequence and among young white dwarves. Finally, we compare our results to the analytic estimate for the spread in UWE, derived from a $\chi$-distribution moderated by the number of observations. These agree with our inferred population limits, though they suggest that we may be biased by a small number of poorly sampled systems. In regions of the sky that are more frequently observed, lower limits could be employed, potentially bringing even more binaries above the threshold for detectability.
\end{abstract}




\section{Introduction}
The \textit{Gaia} space mission provides positions and motions for billions of stars, creating one of the largest stellar catalogues available. Equipped with two telescopes and a combination of an astrometric instrument, a photometer, and a spectrometer, \textit{Gaia} represents a major advance in scanning space astrometry following the \textit{Hipparcos} mission \citep{Perryman1989}. The third data release (DR3, \citealt{Gaia2016, Gaia2023}) extends the observational baseline of point sources from 22 months in DR2 to 34 months. Although the mission collected astrometric data over 10 years, the current data release includes less than half of the full dataset, corresponding to roughly 3 years of observations (and 5 years with DR4). 

\textit{Gaia} follows a predefined scanning law inherited from \textit{Hipparcos}, spinning every 6 hours with its spin-axis tilted $45^\circ$ from the Sun and precessing over $\sim63$ days. The scanning law guarantees that each sky position is observed every 6 months in at least two distinct scan directions, ensuring full-sky coverage. This motion, combined with two astrometric fields of view separated by $106.5^\circ$ (corresponding to a time delay of $\rm 106.5^m$). 


The observation window in the third data release (DR3) spans up to $3$ years, compared to the mission's total $10$-year duration. Binary stars and other companions (planets, brown dwarfs, white dwarfs, neutron stars, and black holes) are all included in the catalogue. Still, only a small fraction of sources (hundreds of thousands out of two billion) have been examined for companions in DR3. Unidentified companions introduce additional noise to single-star astrometric solutions.
At times, the presence of a companion can be a valuable science case to some, but an inconvenience to others. Since binaries are ubiquitous, understanding their behaviour is crucial.

Binary systems consist of two stars orbiting their common centre of mass, with the brighter star as the primary and the dimmer as the companion. When two stars are sufficiently close together, they appear as a single point source dominated by the primary's luminosity and colour. The companion's features are entirely obscured, making the system indistinguishable from a single star. Subtle deviations from single-star motion become the main discriminator, providing valuable insights into their dynamics and evolution \citep{Shu1987, Bate1995}.
In \citealt{P20}, they model the photocentre motion around a binary's centre of mass and show these deviations. Binaries are particularly inconsistent with the 5-parameter model, which includes the on-sky position at a reference epoch, proper motion, and parallax. The deviation is measured by the Unit Weight Error (UWE), which is the square root of a reduced-$\rm \chi^2$.

In this paper, we aim to calculate nominal thresholds for good astrometric fits to differentiate binary system candidates from both single stars and spurious astrometric solutions. In Section \ref{methods}, we lay out the theoretical framework, introduce the definition of $\rm UWE$, and describe the simulation setup based on the \textit{Gaia Universe Model Snapshot}. Section \ref{results} presents the results of our analysis. We first focus on the $\rm UWE$ distributions of single and binary stars across different data releases. Later, we derive the nominal thresholds for DR4 and DR5, confirming earlier estimations as well. Finally, we assess binary detectability as a function of period and other binary parameters. In Section \ref{sec:analytic}, we provide an analytic comparison of our estimations. We summarise the main findings and their implications for future binary detection in Section \ref{concl}, and discuss both the limitations and outlooks for future studies.

\section{Methods}\label{methods}
We simulate binary systems taken from the \textit{Gaia Universe Model Snapshot} (GUMS). GUMS provides a synthetic Milky Way including 3D positions, motions, and a realistic distribution of single and multiple systems. The catalogue represents objects \textit{Gaia} could observe down to $\rm G\simeq20$ at a given epoch\footnote{Instrument effects and observational errors are not included}. Stellar populations are generated using the Besançon Galaxy Model (BGM), with multiple systems produced according to a probability distribution that increases with the mass of the primary. The population is dominated by main-sequence stars, with small contributions from giants, subgiants, and white dwarfs. We do not include the simulated exoplanets in this work. Overall, the majority of stars belong to multiple systems ($68\%$), including binaries and higher-order multiples \citep{Robin2012}.

We model the binary astrometric tracks using the \texttt{astromet} package\footnote{\url{https://github.com/zpenoyre/astromet.py}} \citep{P22a}. \texttt{astromet} closely reproduces \textit{Gaia}'s astrometric fitting pipeline and returns synthetic observables to match the \texttt{gaia\_source} catalogue of single-body astrometric solutions.

The times and angles at which \textit{Gaia}'s field of view passed over are called the \textit{scanning law}. We adopt the Nominal Scanning Law (NSL) \citep{Gaia2016} for the full mission, spanning 10 years of data, corresponding to the final DR5 catalogue. We use the \texttt{gaiascanlaw} package\footnote{\url{https://github.com/zpenoyre/gaiascanlaw}}, which returns, as a function of on-sky position, the epochs (in decimal years), the barycentric coordinate time, and scanning angles (in radians) for each source. The scanning law adopted is based on the \href{https://gaia.esac.esa.int/gost/}{Gaia Observing Schedule Tool} (GOST), which is designed to address degeneracies between parallax and spacecraft motion. We also test shorter baselines corresponding to the upcoming 66-month DR4 release, as well as the 34-month DR3, 22-month DR2, and 14-month DR1 catalogues. Each data release contains all previous data and extends it, and thus, we can expect well-behaved systems to become progressively more precisely constrained. In contrast, sources poorly described by a single-star model (for example, binaries) will exhibit increasing deviations as the timespan of data increases. 

GUMS provides the parameters needed to synthesise the astrometric tracks\footnote{We note that we had to convert the barycentric semi-major axis, $\rm a_{b}$, as recorded in GUMS, to the true semi-major axis, $\rm a$, via $\rm a=a_b\frac{M_A+M_B}{M_A}$}, including orbit orientation angles, masses and magnitudes. We limit ourselves to binaries within 200 pc (local neighbourhood, $\rm G<20$), excluding dim sources beyond which astrometric uncertainties increase rapidly (see Appendix A, \citealt{Lind2021}). The \textit{Gaia} window function changes around $G\sim13$, with brighter sources having 2D positional information, whereas dimmer systems are treated effectively in 1D. However, we do not model this transition explicitly, as we only use the along-scan measurements for fitting, which are substantially more precise than those in the across-scan direction. Including the latter would therefore have only a minor impact in most cases (see \citealt{P20}). We include only binaries with maximum angular separation below 180 mas, beyond which they may be resolved as two separate sources, which excludes long-period systems from our sample.

For a binary system, we label the brighter component as the primary (A) and its companion as the secondary (B). The mass and the light ratios are defined as
\begin{equation}
\begin{aligned}
 q &= \frac{M_B}{M_A} &\quad M_A + M_B &= M_A(1+q), \\
 l &= \frac{L_B}{L_A} &\quad L_A + L_B &= L_A(1+l).
\end{aligned}
\end{equation}
While $l<1$ by definition, $q$ may be greater or smaller than 1. Single stars are simulated by setting the mass and light ratios to zero, eliminating any excess astrometric motion. 

An unresolved binary appears at the position of its centre of light on the sky plane. Because the photocentre and the centre of mass do not coincide, the former orbits around the latter, making the apparent motion of the system non-inertial. Defining $R_l=rl/(1+l)$ and $R_q=rq/(1+q)$ as the distances of the primary from the centre of light and centre of mass, respectively, the displacement of the apparent position from the barycentre is $\Delta \cdot r$ where $r$ is the projected separation and
\begin{equation}\label{Dql}
    \Delta=\frac{R_l-R_q}{r}=\frac{l-q}{(1+q)(1+l)}
\end{equation} \citep{P20}.\\


\subsection{Number of observations in \textit{Gaia} data releases}
\begin{figure*}
    \centering
    \includegraphics[width=\textwidth]{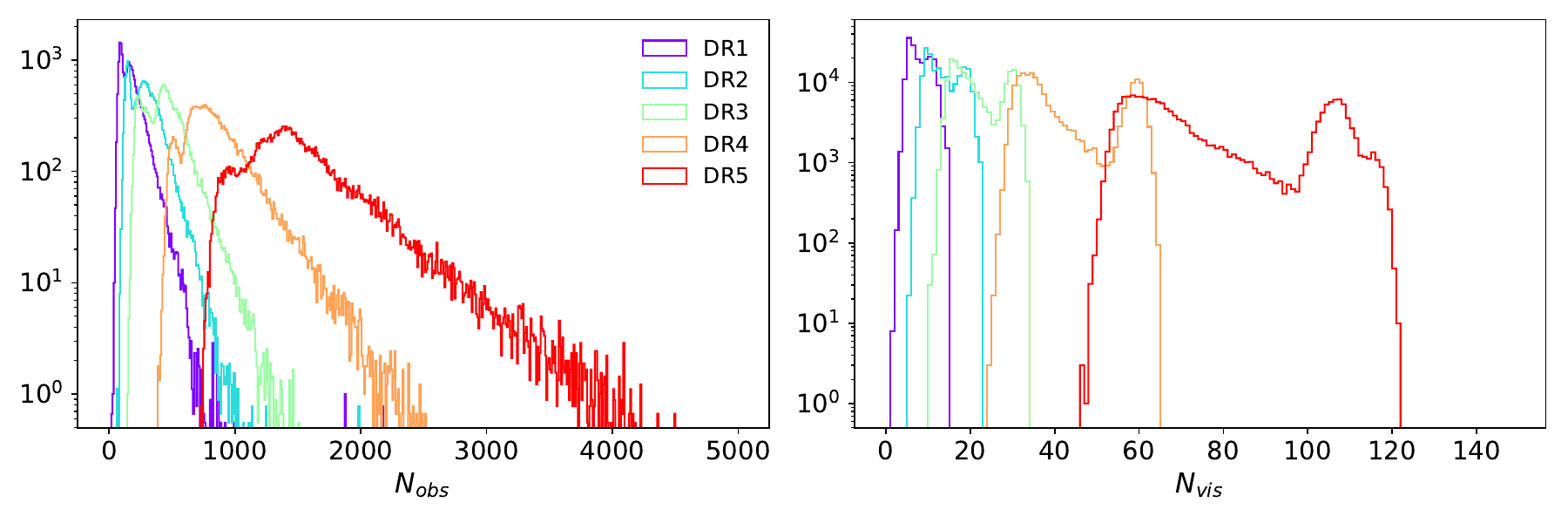}
    \caption{Left: The distribution of the total number of observations per system, $N_\mathrm{obs}$, in each data release for GUMS sources within 200 pc. Right: the corresponding distribution of the number of visibility periods, $N_\mathrm{vis}$, (number of transits separated by more than X hours). Both histograms show a bimodality, mainly due to the ecliptic pole regions ($\beta<-45$° and $\beta>45$°) being scanned with more visibility periods \citep{Lind2018}.}
    \label{Nvorb}
\end{figure*}

The total number of observations per system, $N_\mathrm{obs}$, refers to the number of individual CCD observations. We can also quantify $N_\mathrm{vis}$ to be the number of transits with a minimum separation of four days, a proxy for the number of distinct epochs at which the system is observed \citep{Lind2021}. Figure \ref{Nvorb} shows the distribution of the number of observations and visibility period per system. In principle, each of the two fields of view records 9 CCD observations per transit. However, a minority of observations are lost because our estimate does not account for the fact that only 62, rather 63, CCDs in the focal plane collect astrometric data, and also due to gaps in the data transmission \citep{Lind2021} or observations being rejected as outliers (as processed by the AGIS astrometric pipeline, \cite{Lind2012}). These are not explicitly modelled here, and thus our estimated $N_\mathrm{obs}$ likely represents a slight overestimate. 

As we can see, the total number of observations, $N_\mathrm{obs}$, as well as the number of visibility periods, $N_\mathrm{vis}$, increases over time, resulting in a gradually broader distribution with the peak shifting towards higher values. The spread in the number of observations also increases from the extended baseline, as can be seen when comparing early to late data releases.

Note that, as data obtained from previous releases is included in subsequent releases, solutions do not represent independent measurements of a system over time.

\subsection{Binarity and Unit Weight Error}
With \texttt{astromet}, we fitted a single-star model to each binary system to assess the systematic errors in the astrometry. In particular, we make use of the Re-normalised Unit Weight Error (RUWE), a powerful indicator of stellar multiplicity \citep{P20,B20,P22a,Castro2024} as reported for every source in the \textit{Gaia} catalogue.

We use the observation times, angles, and orbital parameters of the system to compute, via \texttt{astromet}, the positions of the primary ($\alpha_\mathrm{A},\delta_\mathrm{A}$) and the secondary ($\alpha_\mathrm{B},\delta_\mathrm{B}$) in the sky plane. From these, we calculate the coordinates of the centre of light ($\alpha,\delta$). We then express local deviations with respect to some reference position $(\alpha_0,\delta_0)$ as Cartesian positional offsets, $\Delta \alpha^*=(\alpha-\alpha_0)\cos(\delta_0)$ and $\Delta \delta=\delta-\delta_0$.

As \textit{Gaia} provides much more accurate detections along the scanning direction \citep{Gaia2016}, we model all observations as effectively one-dimensional and perform the fits using only the along-scan projected position:
\begin{equation}
    x=\Delta\alpha^* \sin\psi+\Delta\delta \cos\psi+\mathcal{N}[0,\sigma_\mathrm{ast}],
\end{equation}
where $\psi$ is the scanning angle and $\mathcal{N}[0,\sigma_\mathrm{ast}]$ is a normal distribution representing the uncertainty of each measurement. The astrometric error, $\sigma_\mathrm{ast}$, strongly depends on the apparent magnitude $\rm m_G$ for sources dimmer than an apparent \textit{Gaia} magnitude of $\sim 13$ \citep{Lind2021}. By considering a set of observations for a source with measured on-sky positions $x_\mathrm{i}(t_\mathrm{i})$ and comparing them with a 5-parameter single-body fit ($\bar{x_\mathrm{i}}(t_\mathrm{i})$), the $\chi^2$ value is calculated as
\begin{equation}
    \chi^2=\sum_\mathrm{i=0}^{N_\mathrm{obs}}\frac{(x_\mathrm{i}-\bar{x_\mathrm{i}}(t_\mathrm{i}))^2}{\sigma_\mathrm{ast}^2}.
\end{equation}
The expected value for a well-behaved fit should be close to the number of degrees of freedom, i.e., $N_\mathrm{obs}-5$ for a 5-parameter model. The corresponding Unit Weight Error (UWE) is
\begin{equation}
    \mathrm{UWE}=\sqrt{\frac{\chi^2}{N_\mathrm{obs}-5}}.
    \label{eq:UWE_def}
\end{equation}

Three scenarios can be distinguished based on the $\rm UWE$ value. When $\rm UWE\sim1$, the system behaves consistently with the computed single-body model. Values of $\rm UWE<1$ may indicate an overestimation of the astrometric error. $\rm UWE$ values significantly exceeding 1 may either suggest a potential underestimation of $\sigma_\mathrm{ast}$, or the presence of one or more objects within the system contributing to the observed error (e.g., binary motion). Although non-Keplerian perturbations, such as microlensing, may represent potential sources of error, they are not considered here.

A renormalization step is necessary for the real data because the true error, $\sigma_\mathrm{ast}$, is unknown \citep{LindNote2018}. However, as this is a direct input to our astrometric modelling, we report the exact UWE (with no need for renormalization) throughout the rest of this work.

\subsection{Example astrometric tracks}
\begin{figure*}
    \centering
    \includegraphics[width=\linewidth]{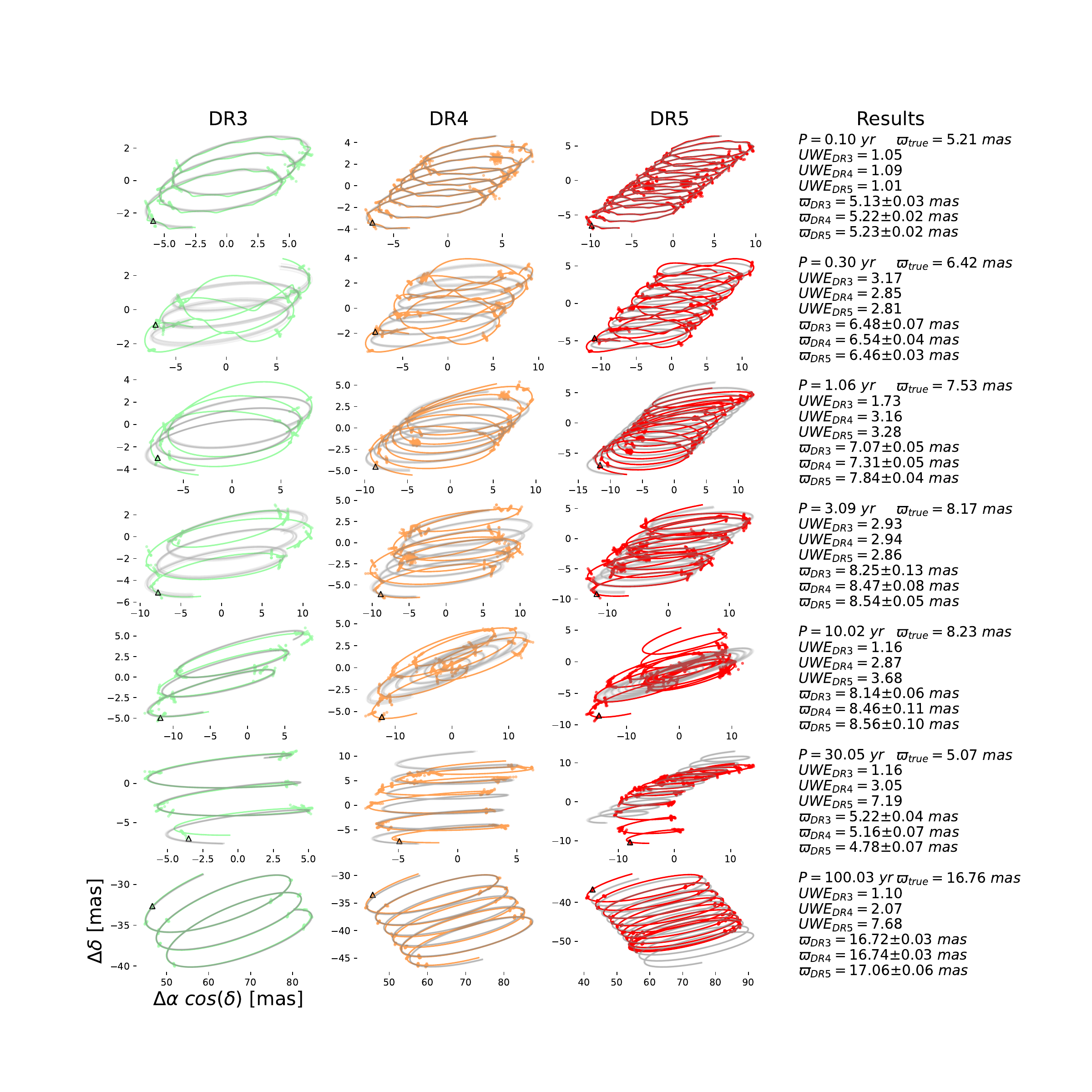}
    \caption{Example astrometric tracks and fits for seven different binaries from GUMS, of increasing period. Simulated data and fits are shown for DR3 (green), DR4 (orange), and DR5 (red). The coloured curves trace the true photocentre motion, points mark the times of observations (9 CCD observations distributed along the scan direction by the scanning law), and black lines represent ensembles of single-body tracks drawn from the distribution of fitted parameters. Triangular markers indicate the first \textit{Gaia} observation of each system. The rightmost column reports the true period and parallax of the fit, along with the fitted UWE and parallax.}
    \label{fig:tracks}
\end{figure*}

We show simulated astrometric tracks of GUMS systems in Figure \ref{fig:tracks}, selected by period. Simulations were performed for the last three data releases. For each source, we overplot a single-body fit to the motion of the centre of light. 

The introduction of a companion adds an extra component of orbital motion, on top of the single-body parallax and proper-motion effect.

In short-period binaries ($P=0.1$ yr and $P=0.3$ yr) with small semi-major axes, the synthetic stellar tracks exhibit small wobbles from the single-body track and represent a source of excess noise to the inferred parameters ($a$). 

Systems with $P\geq1$ yr produce tracks that appear roughly elliptical, but no longer aligned with the parallax ellipse. Now the inferred solution begins to deviate from the true astrometry.

At long periods ($1\lesssim P \ [yr]\lesssim 30$),  orbital motion becomes increasingly irregular and significantly diverges from single-body solutions (most apparent for longer observational windows). For example, intermediate-period binaries ($P=10$ yr, and $P=30$ yr) produce more intricate tracks than can be captured by a single-star model. 

At very long periods, for example, the $P=100$ yr binary shown, the motion may appear consistent with single-body motion. However, we can see that the apparent proper motion (compared to the top row) is very different. Over short timescales, the binary motion appears approximately linear, which biases the inferred fit without causing significant error. For longer observation baselines, the single-body fit starts to deviate, and the deviations from a true single-body motion become apparent.

In conclusion, many systems in our sample appear consistent with a single-star model in earlier data releases (particularly those with $P=0.1, 10, 30,$ and $100$ yr). As the observational baseline broadens, however, the underlying orbital complexity becomes clear, revealing that earlier releases can yield misleading classifications of binarity.

\section{Results}\label{results}
\begin{figure}
    \centering
    \includegraphics[width=\columnwidth]{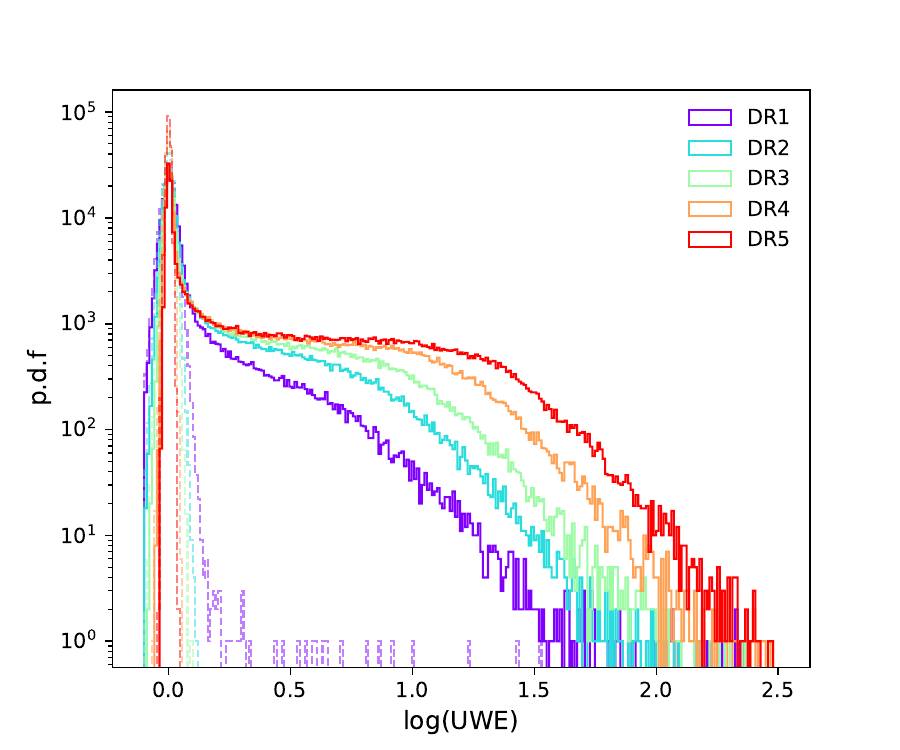}
    \caption{The distribution of simulated $\rm UWE$ for single (dashed) and binary (solid) stars, as shown for each data release.}
    \label{fig:pdf}
\end{figure}

For each data release, the UWE was computed as shown in Section \ref{methods}. We examine the probability distribution function (p.d.f) of UWE shown in Figure \ref{fig:pdf}.

The binary distribution extends to higher $\rm UWE$ values, increasing over time. This trend reflects the ability of a longer baseline to capture larger deviations and enable more accurate binary identification. In contrast, single stars exhibit progressively sharper distributions over time, with DR1 displaying a much broader distribution than DR5. 

Both distributions peak at $\mathrm{UWE}\sim1$; thus, no UWE criterion can separate all binaries from single stars. Systems with small or very long binary orbits, or inconveniently aligned systems, will always be considered single stars \citep{P20,P22a}.

\subsection{Nominal criteria for good astrometric fits}\label{sec:thresholds}
Now we calculate the UWE limits, $\rm UWE_{DRN,lim}$, beyond which fits can be deemed inconsistent with single star solutions, for each data release. 

We do this by first calculating the survival function (s.f., equal to one minus the cumulative distribution function) of the single star fits. We then ask when the s.f. falls beneath some probability cut-off value, here $10^{-6}$ corresponding to one in a million single stars being excluded as bad fits. 

\begin{figure}
    \centering
    \includegraphics[width=\columnwidth]{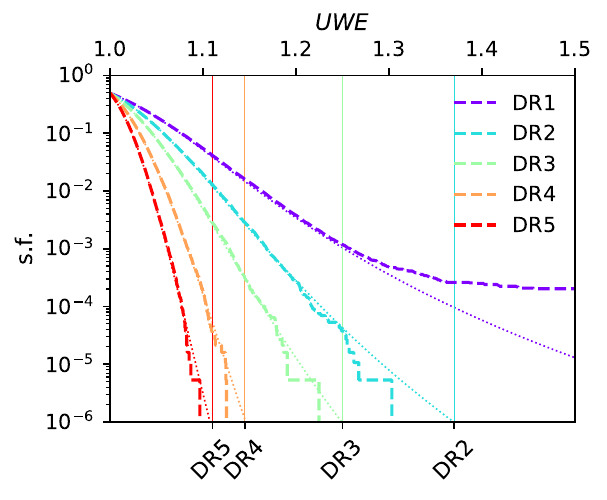}
    \caption{The fraction of systems below a given $\rm UWE$ is shown for single stars (dashed lines). We fit each with a Student's-$t$ distribution (dotted line) to determine the approximate upper limit on $\rm UWE$ (vertical solid lines), corresponding to a threshold that would exclude fewer than one in a million well-behaved single stars.}
    \label{fig:sdf}
\end{figure}

As the threshold value is very low, we extrapolate the simulated distribution by fitting a Student's-$t$ distribution to the UWE distributions of single stars. The fits were performed with \texttt{scipy.stats.t}, of the form
\begin{equation}
    f(x, \nu) = \frac{\Gamma\left(\frac{\nu + 1}{2}\right)}{\sqrt{\nu \pi} \, \Gamma\left(\frac{\nu}{2}\right)} \left(1 + \frac{x^2}{\nu}\right)^{-\frac{\nu + 1}{2}}
    \label{studt}
\end{equation}
where $\nu$ represents the degrees of freedom, $\Gamma(n)=(n-1)!$ is the Gamma function, and $x=\frac{\mathrm{UWE}-\mu}{\sigma}$. A Student's-$t$ distribution provides a more appropriate description than a Gaussian because of its broader wings. The distributions and fits can be seen in Figure \ref{fig:sdf}.

As expected, the threshold differs across data releases, decreasing over time as the observational baseline increases and more measurements become available. We note that the UWE distribution is not uniform across the sky. Regions with poorer scanning-law coverage tend to exhibit larger astrometric uncertainties and increased parallax errors. These effects are most prominent near the ecliptic plane compared to the ecliptic poles. We explore the difference between the astrometric behaviour near the ecliptic and poles in Appendix \ref{app: eclpol}.

For DR2 and DR3, the derived thresholds are consistent with previous studies: $\mathrm{UWE_{DR2,lim}}=1.37$ ($1.4$ in \citealt{Lind2018}) and $\mathrm{UWE_{DR3, lim}}=1.25$ \citep{P22a,Castro2024}. For DR1, however, $\rm UWE$ can be arbitrarily large, due to a small number of observations, and thus we do not present a limiting value.

For the upcoming data releases, we obtain $\mathrm{UWE_{DR4,lim}}=1.15$ and $\mathrm{UWE_{DR5,lim}}=1.11$. These values are consistent with the existing data and provide thresholds for identifying additional binaries in DR4 and DR5. Although the observational window doubles from DR4 to DR5, the relative change in the threshold is smaller. Thus, longer baselines will continue to improve the accuracy of astrometric fits, increasing the number of detectable binaries.

The thresholds we derived here are summarised in Table \ref{tab:thresholds}, along with the duration (in months) of each data release.

\begin{center}
\begin{tabular}{ c c c }\label{tab:thresholds}
 Data Release & Length (months) & $\rm UWE_{lim}$ \\ 
 \hline
 1 & 14 & - \\  
 2 & 22 & 1.37 \\     
 3 & 34 & 1.25 \\     
 4 & 66 & 1.15 \\     
 5 & 126 & 1.11 \\     
\end{tabular}
\end{center}


\subsection{Binary detectability as a function of period}\label{sec:bindet}
With these UWE thresholds, we can check which simulated binaries would be detectable by \textit{Gaia}. 

\begin{figure*}
    \centering
    \includegraphics[width=0.85\linewidth]{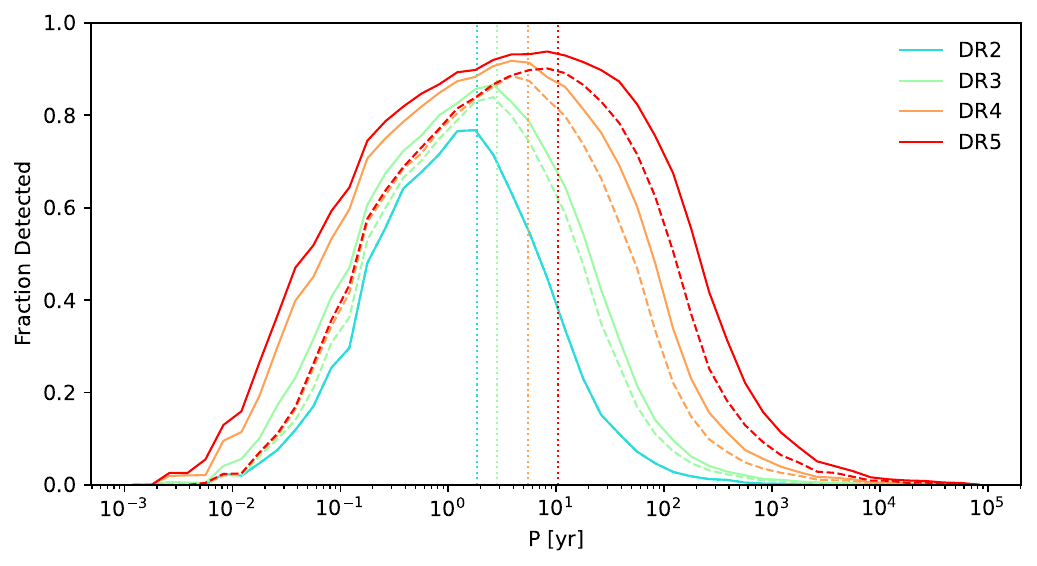}
    \caption{The fraction of detectable binaries as a function of period. A system is considered detectable when $\mathrm{UWE_{DRN}} > \mathrm{UWE_{DRN,limit}}$ (solid lines). The same set of curves has also been plotted using a fixed threshold of 1.4 (dashed lines). Vertical lines show the observational baseline corresponding to each data release.}
    \label{fig:fd}
\end{figure*}

Figure \ref{fig:fd} shows the fraction of detected binary systems as a function of period. This fraction was computed by applying the condition used to distinguish binaries from single stars: $\mathrm{UWE_{DRN}}>\mathrm{UWE_{DRN, limit}}$ with $\rm N=[2,3,4,5]$.

Each distribution exhibits a peak in period, ranging from $P_\mathrm{max, DR2}\simeq1$ yr to $P_\mathrm{max, DR5}\simeq10$ yr, corresponding closely to the baseline of each survey. Longer baselines enable \textit{Gaia} to detect more binaries. The small deviations caused by short-period binaries increase in significance when observed over multiple orbits. Long-period systems may display significant curvature with an extended baseline, as now a good fraction of a whole orbit is observed, differentiating the orbital motion from proper motion.

We see that the fraction of long-period binaries detected increases by around 10-20\% with each subsequent release. The detected fraction of short-period binaries increases by a more modest 5-10\%. While the exact fractions depend on the specific sample of simulated binaries considered, the overall increase in detectability is expected to hold more generally.

The choice of an appropriate $\rm UWE_{DRN,lim}$ to filter and select binary systems detected by \textit{Gaia} is essential. For comparison, we computed the binary fraction using a uniform cut of $\mathrm{UWE_{DRN}}>\mathrm{UWE_{DR2, limit}}=1.4$, i.e., applying DR2 conditions to all data releases (Figure \ref{fig:fd}). 

Using this threshold, the detectable fraction of longer-period systems is slightly reduced. However, shorter-period binaries are much less likely to be detected, with almost no increase from DR4 to DR5. This is because, for systems where we observe multiple orbital periods, the UWE asymptotes to a constant value (as predicted in \citealt{P20}). As precision improves and the UWE limit decreases, this fixed UWE becomes increasingly significant over time. Thus, adopting the same threshold across all data releases causes a significant loss of detectable binaries (particularly at low mass ratios, as shown by \citealt{Elbadry2024}).

It is interesting to note that the detected fractions remain non-zero at both long and short periods, suggesting a small fraction of ideal binaries can still be detected by UWE at extreme periods. We will explore these systems in the next sections.

\subsection{Binary detectability as a function of other binary properties}\label{sec:fd_orbPe}
The binary period is the single most important parameter for determining astrometric detectability, but other parameters also modify which binaries are likely to exhibit detectable deviations from single-body motion.

From DR3 onwards, we show the distribution of the fraction of detected sources as a function of $P$, plotting it against different parameters such as eccentricity ($e$), mass ($q$), and light ratio ($ l$).

\begin{figure*}
    \centering
    \includegraphics[width=\textwidth]{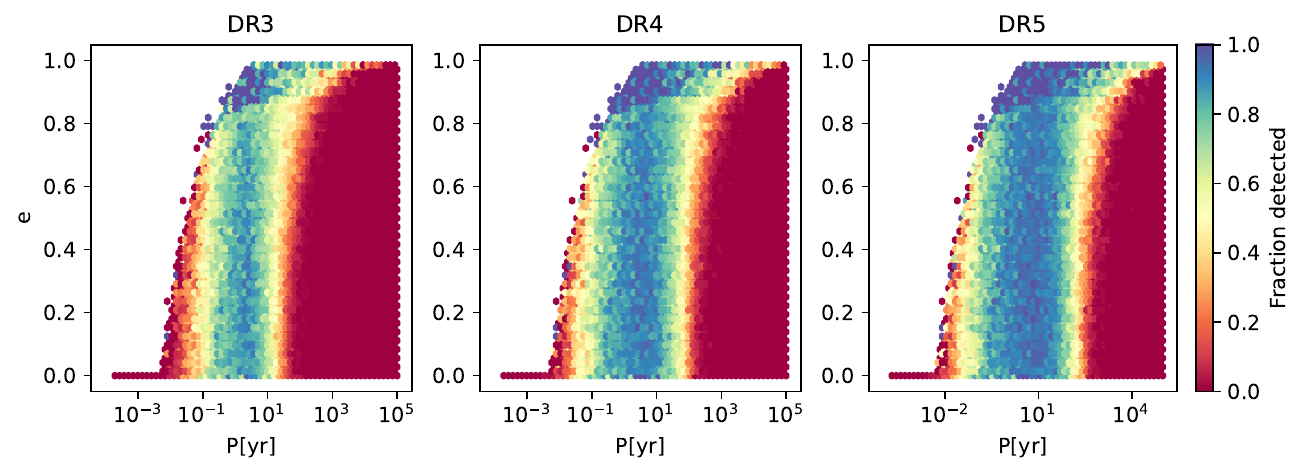}
    \caption{Similar to Figure \ref{fig:fd}, we present the binary detectability as a function of period and eccentricity for DR3 (left panel), DR4 (middle panel), and DR5 (right panel).}
    \label{fig:Pvecc_drs}
\end{figure*}

Starting from Figure \ref{fig:Pvecc_drs}, the grid displays the detected fraction in period-eccentricity space. Tidal circularization (as prescribed by the GUMS population synthesis) limits the eccentricity of short-period systems. At low eccentricity ($e\lesssim 0.5$), the detection fraction is essentially independent of $e$. The detection fraction reduces significantly beyond 100 years (in DR4 and DR5).

There is also a significant population of detectable high-eccentricity systems at large periods. These correspond to orbits where the fast periapse passage of the binary falls within the observing window. Even if we see only a fraction of an orbit, the motion near periapse represents short-timescale changes in speed and direction, which cannot be captured by a single-body fit. Thus we can see that the long tail of high-period detectable binaries described in the previous section consists almost entirely of highly eccentric ($e\gtrsim0.8$) systems.

Also, the distributions show a sharp, step-like increase in the fraction of detected binaries with moderate periods and $e \gtrsim 0.9$. This feature reflects the random generation of eccentricities in the GUMS model and arises because only (bright) massive stars with lower mass companions populate this region. These objects are ideal astrometric candidates due to the low astrometric error even at the edge of our 200 pc volume\footnote{Private communication from DPAC expert C. Babusiaux.}.

\begin{figure*}
    \centering
    \includegraphics[width=\textwidth]{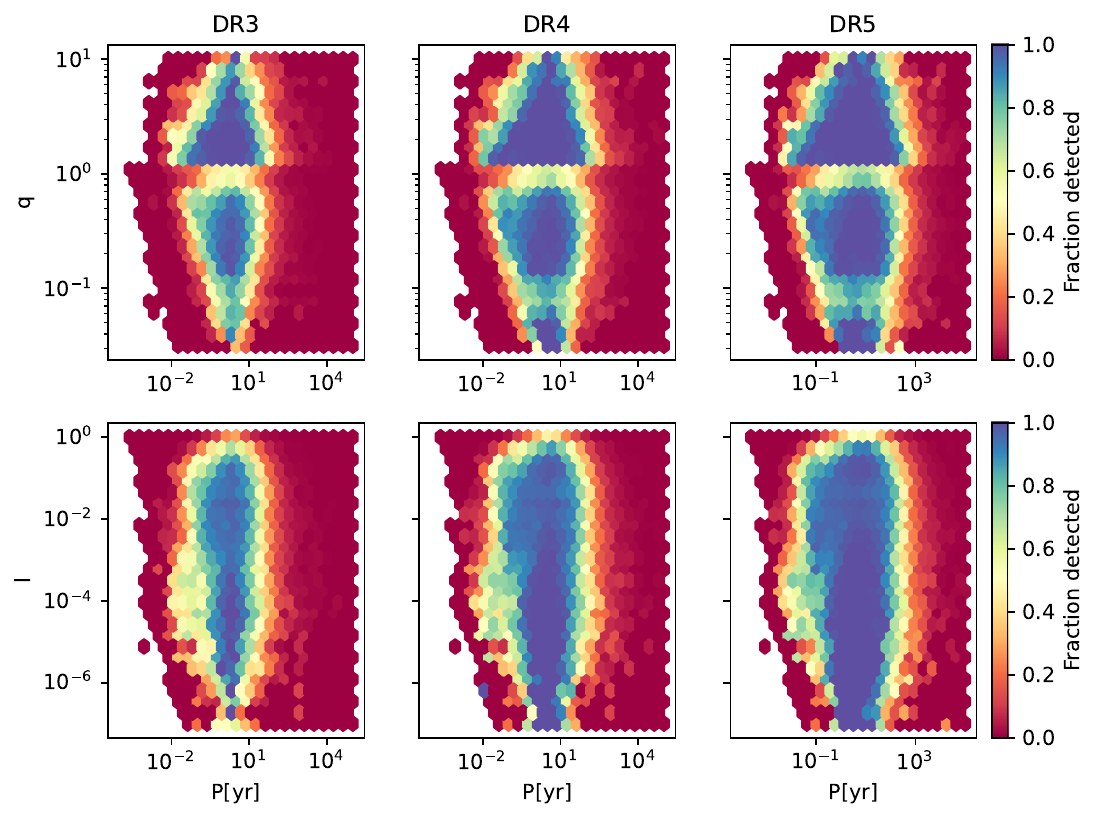}
    \caption{Binary detectability as a function of mass ($q$, top row) and light ($l$, bottom row) ratio for DR3, DR4, and DR5.}
    \label{fig:Pvql}
\end{figure*}

Similarly, we plot the detectability as a function of orbital period and mass ratio $q$, and light ratio $l$, in Figure \ref{fig:Pvql}. The fractional offset between the centre of mass and the photocentre, $\Delta$, is proportional to $q$-$l$ (equation \ref{Dql}) and thus twin systems, where $q\sim l\sim 1$, show decreased detectability. 

For most systems, the light ratio is much smaller than the mass ratio, and thus $\Delta\approx \frac{q}{1+q}$ and the detectability is relatively insensitive to $l$ for $l \lesssim 0.1$. The amplitude of the astrometric signal decreases with decreasing $q$, and thus the detectable fraction reduces at low $q$. As we do not consider GUMS exoplanets (with $q \lesssim 10^{-3}$), our sample only extends to $q\sim 0.03$, but we can see that the detection fraction at these mass ratios is much reduced and confined to periods close to the observing time of each data release. Periods significantly shorter or longer than the observing baseline produce astrometric signals that are undersampled or incomplete, reducing detectability. Very low mass ratio systems (such as exoplanets) will likely be detectable only in our direct vicinity, where the large $\varpi$ can offset the small $q$.


High mass ratio binaries are those with massive non-luminous companions (white dwarfs or compact objects), and whilst these provide excellent astrometric signals, they are relatively rare. Thus, the detectable fraction decreases for increasing $q>1$ as the distance to the nearest example of such a system increases.

In the more ideal parts of parameter space (for example, small $l$, which also suggests a bright primary, and periods of a few years), the detection fraction can reach 100\%. This shows that while the astrometric deviations are modulated by factors, such as the orientation of the binary, they do not significantly dampen the prospects for detection. However, for a sample extending to a larger volume, we would no longer expect all such systems to be detectable, as the astrometric signal is proportional to $\varpi$ and the astrometric error decreases for dimmer sources.


\begin{figure*}
    \centering
    \includegraphics[width=\linewidth]{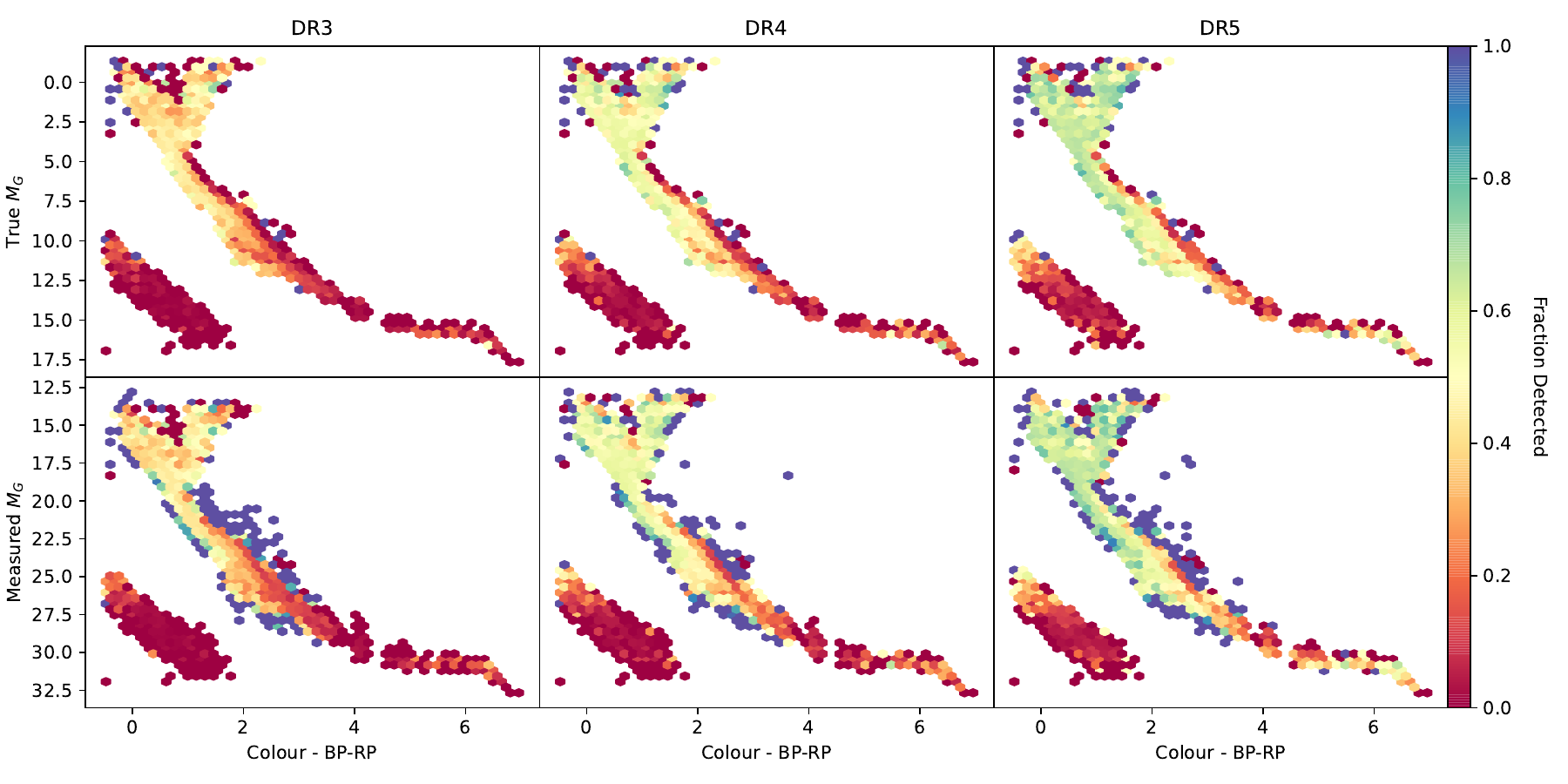}
    \caption{Binary detectability as a function of true (upper panels) and measured (bottom panels) absolute magnitude, $M_\mathrm{G}$, for DR3, DR4, and DR5.}
    \label{fig: HR}
\end{figure*}

Figure \ref{fig: HR} shows the detectability of our sample of binaries across the colour-magnitude diagram. We use both the true absolute magnitudes, based on the true distance (top panels), and the measured absolute magnitudes based on the distances derived from the astrometric fits (bottom panels). \

Generally brighter systems are more likely to be detected. As we move from DR3 to DR5, the fraction of detectable systems increases, rising to over two-thirds for much of the main-sequence. 

A prominent exception is seen with twin binaries (with equal mass and brightness, sitting 0.75 magnitudes above the main sequence) for which $\Delta$ (equation \ref{Dql}) goes to zero. These systems remain largely undetected, because the centre of mass coincides with the photocentre ($q\sim l\sim1$), causing them to appear as bright single stars \citep{Moe2017}. Nevertheless, they may still be identifiable from their excess brightness.

Bright white dwarf binaries also become increasingly detectable. These systems, located at the tip of this region in the diagram, represent a subgroup of hot and young white dwarf binaries. With a 10-year observational baseline, we therefore expect a significant increase in the number of detected white dwarf binaries in future data releases, consistent with early predictions of \textit{Gaia}'s white dwarf yield \citep{Carrasco2014}. As a result, constraints on population and evolution models can be substantially improved.

When we construct the measured absolute magnitudes from the measured parallaxes, we see significant extra scatter. Previously unoccupied regions are generally associated with very high detectability fractions, as a significant astrometric effect is needed to cause these large parallax shifts. The degree of scatter is somewhat reduced in later data releases, but a minority of systems still appear in unphysical regions. We note that from DR4 onwards, the reported astrometric fits will be adjusted if there is detectable orbital motion or curvature, and so may be more accurate than our values fitted with the single star pipeline.
\section{Analytic comparison}\label{sec:analytic}
For a well-behaved single source, we can also predict the expected distribution of UWE values analytically. The UWE is the square root of a reduced $\chi^2$ statistic, and thus the spread of values can be replicated from a renormalised $\chi$-distribution:

\begin{equation}
 p(\mathrm{UWE},\nu)= \frac{(\nu\cdot \mathrm{UWE})^{\nu-1} e^{-(\nu\cdot \mathrm{UWE})^2/2}}{2^{\nu/2}\Gamma(\nu/2)}
\end{equation}
where $\nu$ is the number of degrees of freedom (here assumed to be $N_\mathrm{obs}-5$ for a 5-parameter fit).

\begin{figure}
    \centering
    \includegraphics[width=\columnwidth]{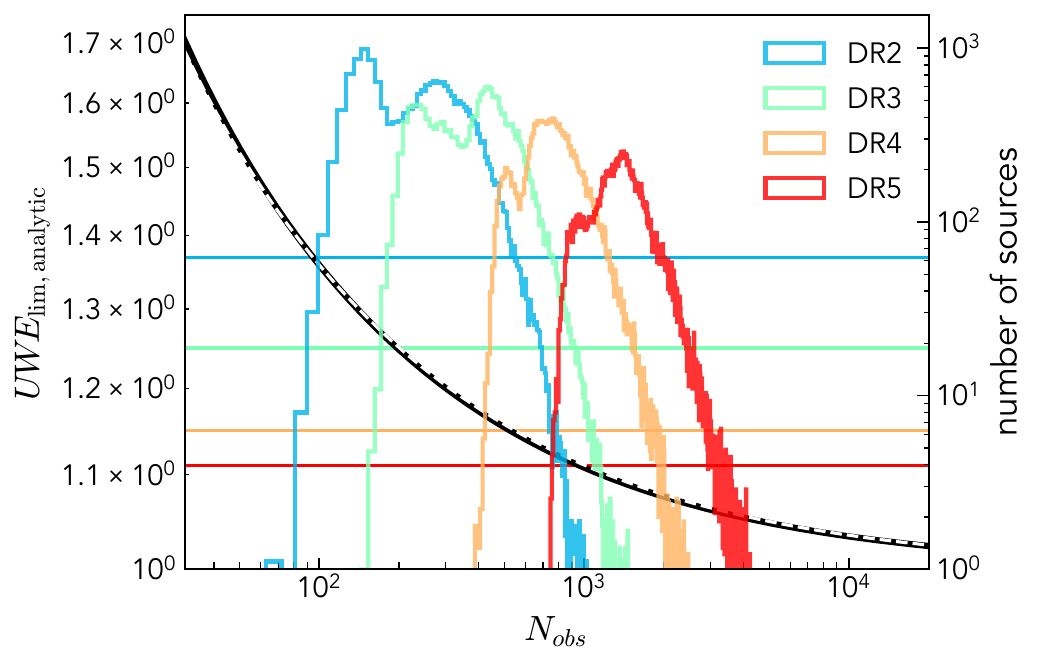}
    \caption{The analytic UWE limit (black line) for single stars observed $N_\mathrm{obs}$ times. For context, we also show the distribution of the number of observations in each data release for our GUMS sources (equivalent to Figure \ref{Nvorb}), along with the suggested limiting UWE as found in Section \ref{results} (horizontal lines). A simple approximation to the limiting $\rm UWE$ (as given by equation \ref{eq:uwelim_approx}) is shown as a dashed white line.}
    \label{fig:chi}
\end{figure}

Figure \ref{fig:chi} shows the expected UWE limit, corresponding to a survival fraction of $10^{-6}$, as a function of $N_\mathrm{obs}$. We can see that our suggested UWE limits correspond to the lowest $N_\mathrm{obs}$ sources in each sample. This suggests either that these dominate the convolved distribution, or that our Student's-$t$ distribution slightly overestimates the occupancy at high UWE. Both suggestions are true: sources with a high $N_\mathrm{obs}$ could be expected to exhibit a significantly lower spread in $\rm UWE$ than the population average in each data release. Since our models include no other sources of noise, we naturally expect this analytic cut-off to be a lower limit when compared to actual data.

Using a single $\rm UWE_{lim}$ for each dataset has a pleasing simplicity, but, as this analysis shows, in future work it may be useful to adjust the limiting value used for systems as a function of $N_\mathrm{obs}$. To facilitate this, we also include a simple approximate form in Figure \ref{fig:chi} of the form
\begin{equation}
\label{eq:uwelim_approx}
\mathrm{UWE_{lim}}(N_{\mathrm{obs}}) \sim 1+\frac{3.5}{\sqrt{N_\mathrm{obs}-5}}.
\end{equation}

\section{Conclusions}
\label{concl}
Unresolved astrometric binaries are the main focus of this study. As they are not spatially resolved, they present a challenge to detect and characterise. Here we explore identification through deviations from a 5-parameter single-body model, which provides an adequate fit for single stars but large errors for binaries with unmodelled orbital motion. Particularly, the orbital perturbation introduces additional astrometric noise, which can be measured through the (R)UWE value. This parameter determines the goodness of fit: values close to unity indicate consistency with the single-star model, whereas significantly higher values suggest that the single-body model is insufficient to fit the data. Given the ubiquity of stellar companions, one of the most common causes of high UWE will be binarity.

This study aimed to establish nominal upper limits on the UWE parameter for good single-body fits in future \textit{Gaia} data releases, and the subsequent detectability of various binary systems. We use simulated binaries from the Gaia Universe Model and investigate the long-term astrometric behaviour of single and binary stars. The \texttt{astromet} package provided the tools required to compute and fit the astrometric tracks, consistent with the \textit{Gaia} data processing and previous studies \citep{B20,P20,P22a}. We have used the full 10-year nominal scanning law to compute the times and angles at which systems are observed, and we can thus see both the improvements and limitations of existing and upcoming data releases.

The UWE distributions for single and binary systems are discussed in Section \ref{sec:thresholds}. In Section \ref{sec:bindet}, we estimated the $\rm UWE$ limits for fits consistent with a single-body in each data release. We do this by analysing the probability distribution of the generated single-star sample. Accordingly, we propose the following thresholds for identifying binary systems in DR4 and DR5: $\rm UWE_{DR4,lim}=1.15$ and $\rm UWE_{DR5,lim}=1.11$, respectively. We recover thresholds for the existing data releases of 1.37 for DR2 (compared to the value of 1.4 suggested by \citealt{Lind2018}) and 1.25 for DR3 (agreeing with \citealt{P22a} and \citealt{Castro2024}).

As the timespan of observations increases, the $\rm UWE$ detection thresholds decrease and more binaries are detectable, primarily as a function of their period. Systems with longer periods benefit from the extended baseline, which increases the likelihood of observing a significant fraction of an orbit. At shorter periods, the UWE tends to a constant, but as the precision of single-star characterisation improves, the significance of the deviations increases. Earlier releases yielded detections only within a limited period range of $0.1-10$ years. By DR5, this region will broaden significantly, with detections possible from $0.01$ to $100$ years and leading to a substantial increase in the number of detectable binaries.

In Section \ref{sec:fd_orbPe} we examined the dependance of the detectability on other binary parameters: the eccentricity, mass ratio, and light ratio. We see that the longest detectable periods correspond to highly eccentric systems where we are lucky enough to observe them near periapse, a probability that increases linearly with the timespan of observations. Low to moderate eccentricity systems are detectable up to about 100-year periods (in DR4 and later). The detection fraction reduces for mass ratios that are very small, very large, or close to one. Small mass ratio systems give only low-amplitude astrometric motion, detectable only if they are very close by. High mass ratio systems are rare, and we must look further and further to the nearest example. The astrometric signal is suppressed if the light ratio and mass ratio are very close, and this is most frequently the case for twin systems with $q\sim l \sim 1$.

Finally, in Section \ref{sec:analytic}, we calculate and compare the nominal analytic UWE limit via a $\chi$-distribution, a direct function of the number of observations (and thus only indirectly dependant on the timespan of the survey). We show that our population-scale UWE limits are biased towards the least frequently observed systems in each data release, as these exhibit the widest spread of possible values. The analytic UWE limits should be considered a hard lower bound, as they include no additional sources of noise. They also suggest an interesting alternative to using a single limit for each data release. More frequently observed systems are expected to be better constrained, and thus, we may have even more capacity to differentiate good and poor astrometric solutions.

The sample of binaries from GUMS allows us to probe the behaviour of a broad and generally representative array of systems, although there are still limitations compared to the full scope of possible binaries and observational complexities. Firstly, GUMS lacks high-density regions (e.g., clusters) and galaxy-scale substructures, where reliable astrometric measurements are made more difficult due to crowding. The simulated populations are primarily restricted to main-sequence stars, giants, and white dwarfs, and lack extremely compact companions such as neutron stars or black holes, rare objects but of great interest, and which are expected to yield strong astrometric signals \citep{Andrew2022}. The sample is limited to $200$ pc and to systems with angular separations below $180$ mas, favouring brighter stars with more accurate astrometry, while more distant populations ($\sim400-500$ pc) and massive binaries remain unexplored. Finally, our analysis adopted a 5-parameter single-star model, and could be extended to 7- (and 9-) parameter solutions, which include two (four) additional acceleration (and jolt) components. This extension will allow more accurate detection of long-period systems \citep{Elbadry2024}. Future work should thus ideally extend the catalogue to larger distances, include more realistic binary populations and substructures, and compare to more complex possible astrometric fits.

Finally, in our simulations, we include only binarity and astrometric noise, neglecting the contribution of tertiary companions, blended sources, misidentified objects, or data gaps. Therefore, the UWE thresholds in this scenario serve only as lower limits. Consequently, our results provide either an ideal limit, if the goal is to include binaries, or a conservative limit otherwise.




\section*{Acknowledgements}
The authors acknowledge the use of the \textit{Gaia Observation Schedule Tool} (\href{https://gaia.esac.esa.int/gost/}{GOST}) maintained by the \textit{Gaia} Data Processing and Analysis Consortium (DPAC). The Gaia mission website is \url{https://www.cosmos.esa.int/web/gaia}. We would also like to thank the referee for their detailed
and significant suggestions and improvements. ZP acknowledges support from European Research Council (ERC) grant number: 101002511/project acronym: VEGA P.

\section*{Data Availability}
The data underlying this article will be shared on reasonable request to the corresponding author.



\bibliographystyle{mnras}
\bibliography{references} 



\appendix\label{app: eclpol}
\section{Ecliptic: Plane and Poles}
We briefly show the different astrometric behaviour near the ecliptic plane ($|\beta|<15^\circ$) and ecliptic poles ($|\beta|>45^\circ$).

\begin{figure*}
    \centering
    \includegraphics[width=\linewidth]{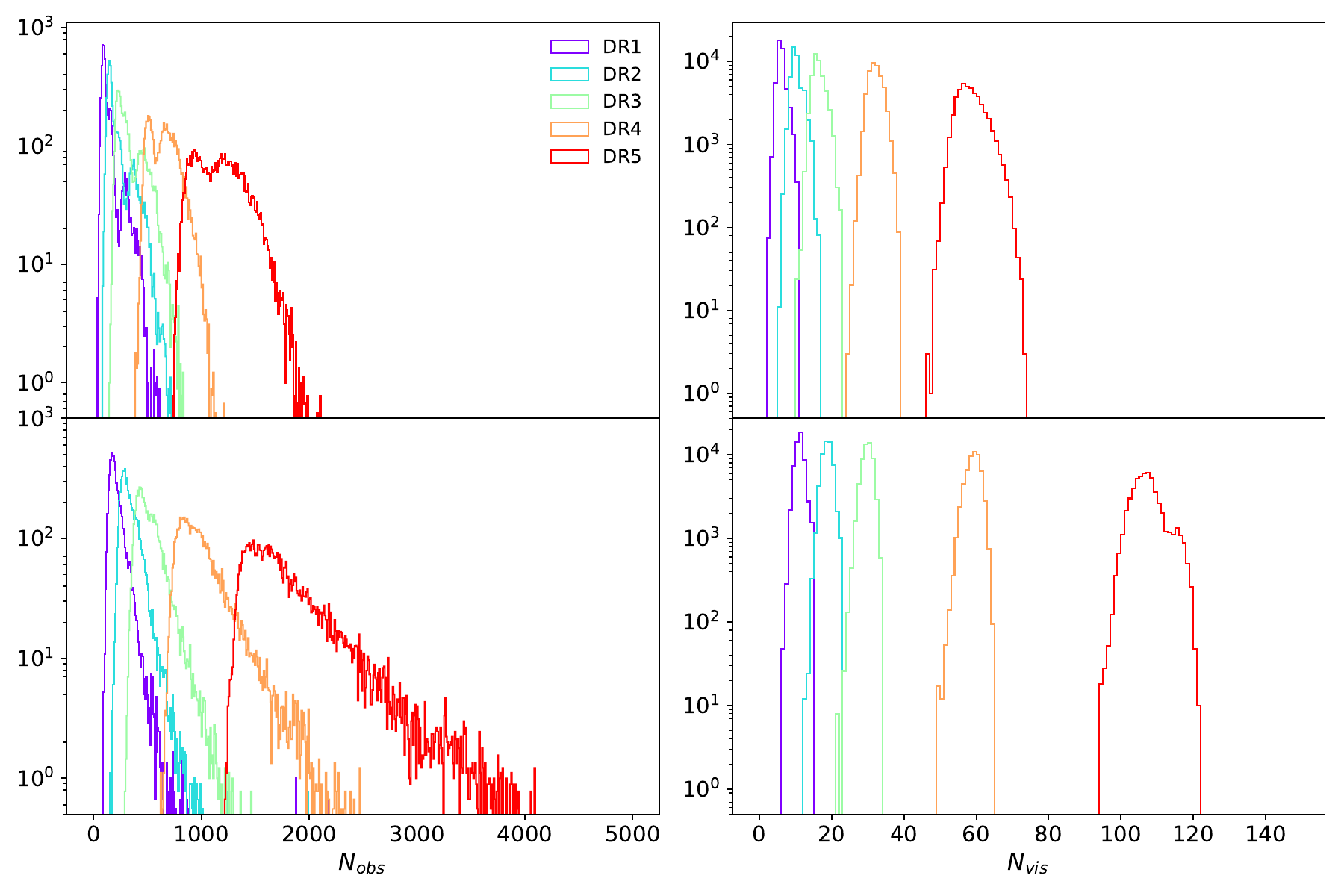}
    \caption{Same as Figure \ref{Nvorb}, but separated by ecliptic latitude, $\beta$. Upper panels show sources near the ecliptical plane, and lower panels near the ecliptical poles (bottom panels).}
    \label{fig: nobs_beta}
\end{figure*}

Figure \ref{fig: nobs_beta} shows that both $N_\mathrm{obs}$ and $N_\mathrm{vis}$ significantly increase toward the ecliptic poles (bottom panels) compared to the ecliptic plane (upper panels), which is scanned less frequently. At the poles, the $N_\mathrm{obs}$ distributions are significantly broader, by nearly 50\%, indicating that the existing systems are better observed at the poles compared to the plane, given the scanning law geometry \citep{Gaia2016}. We also notice a slight shift of the distribution peak towards higher values. This effect is more pronounced for $N_\mathrm{vis}$, for which a stronger dependance on ecliptic latitude is expected given the scanning law geometry \citep{Gaia2016}. 

\begin{figure*}
    \centering
    \includegraphics[width=\linewidth]{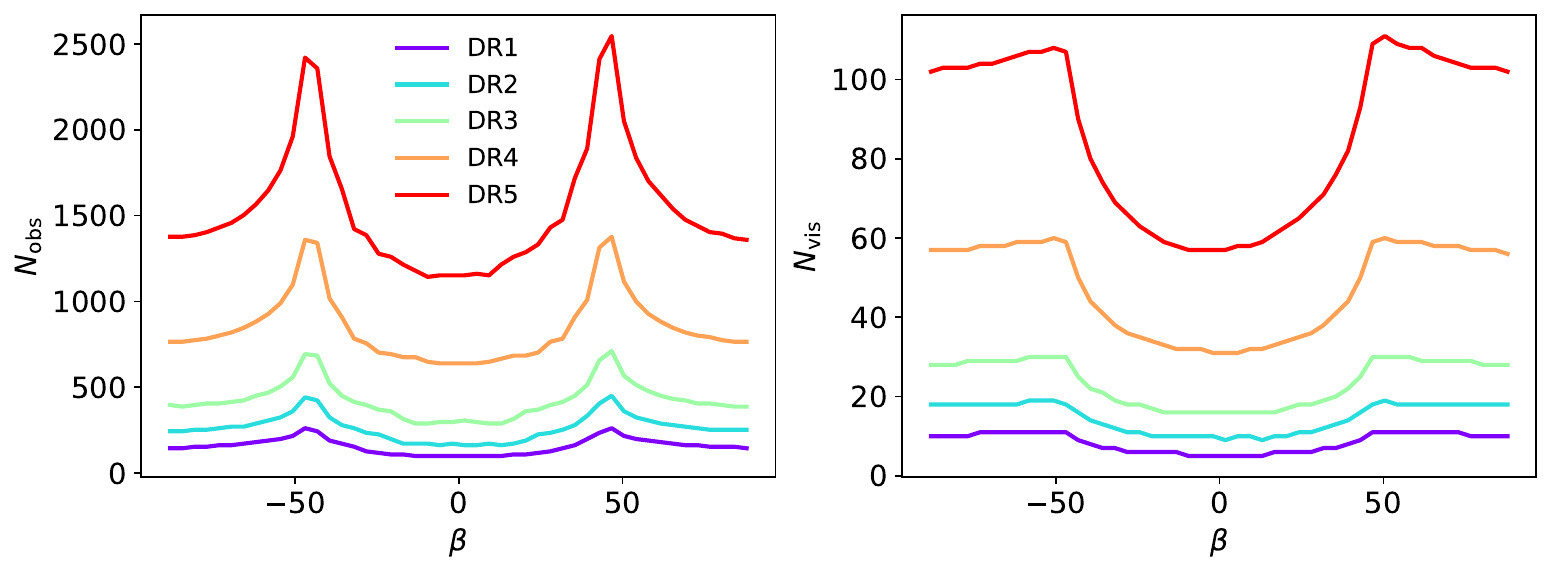}
    \caption{Median distributions for $N_\mathrm{obs}$ (left panel) and $N_\mathrm{vis}$ (right panel) as a function of the ecliptical latitude, $\beta$, shown for each data release.}
    \label{fig: nvis_beta}
\end{figure*}

As shown in Figure \ref{fig: nvis_beta}, both $N_\mathrm{obs}$ (left) and $N_\mathrm{vis}$ (right) peak around $|\beta|=45^\circ$ and reach a minimum near the ecliptic plane ($\beta=0^\circ$). However, their behaviour differs at higher latitudes. While $N_\mathrm{obs}$ accumulates around $|\beta|=45^\circ$ due to the scanning law geometry, as expected \citep{Gaia2016, Holl2023}, $N_\mathrm{vis}$ exhibits a plateau. This difference arises because $N_\mathrm{vis}$ counts only well-separated observations. Near the ecliptic plane many observations are clustered in short time intervals, which increases $N_\mathrm{obs}$ but contributes less to $N_\mathrm{vis}$. In contrast, sources at higher latitudes remain within the scanning path of the spin-axis precession for longer periods, resulting in more evenly distributed observations and therefore maintaining high $N_\mathrm{vis}$ values.   

The median profiles become progressively steeper across data releases, enhancing the differences in scanning-law coverage (see Figures 3 and 4 from \citealt{distefano2023}).
This reflects the increasing sensitivity to $\beta$ as the observational baseline grows, with peak values increasing by roughly an order of magnitude from DR1 to DR5.


\begin{figure*}
    \centering
    \includegraphics[width=\linewidth]{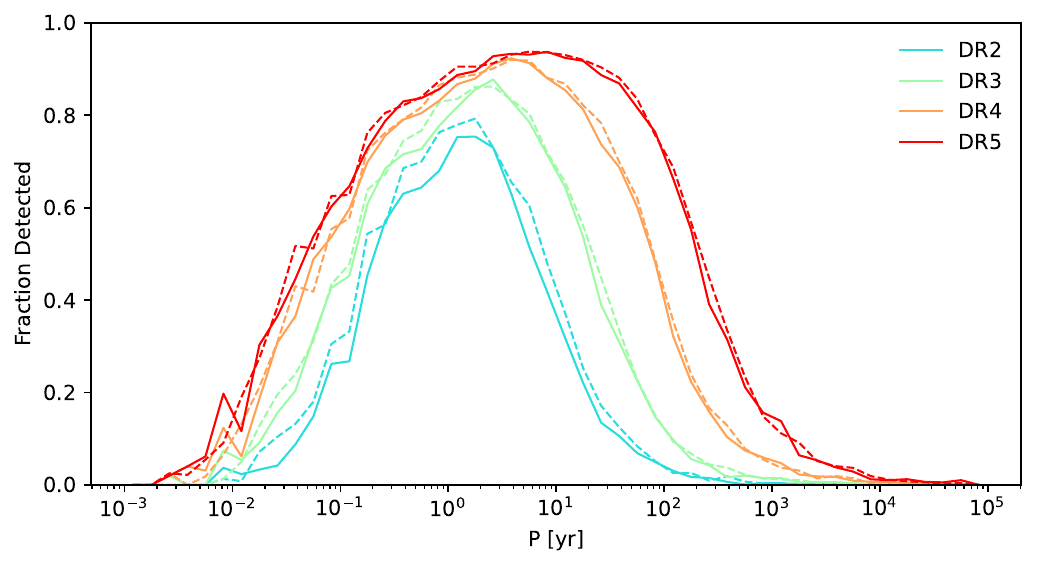}
    \caption{Same as Figure \ref{fig:fd}, showing contributions near the ecliptic plane (solid lines) and near the poles (dashed lines).}
    \label{fig: frac_beta}
\end{figure*}

Figure \ref{fig: frac_beta} shows the effects of $\beta$ on binary detectability. The distributions for sources near the poles (dashed lines) are slightly broader than those in the ecliptic plane (solid lines), indicating that binary detection is enhanced at higher latitudes \citep{Gaia2016}. This is also consistent with $\mathrm{UWE}\sim N_\mathrm{obs}^{-1/2}$, as shown in equation \ref{eq:uwelim_approx}, thereby producing only a small shift in detectability. We note that this discrepancy is more prominent in earlier data releases, especially DR2, due to its shorter baseline (Table \ref{tab:thresholds}). As we approach the full mission 10-year baseline, this feature becomes less significant, as the overall number of transits per source grows significantly, as expected (Figure \ref{fig: nobs_beta}). As a result, the relative impact of the scanning law geometry on detectability is reduced, leading to increasingly similar distributions.


\bsp	
\label{lastpage}
\end{document}